\documentstyle[12pt,aaspp4,epsfig]{article}
\def\spose#1{\hbo}

\def\beq{\begin{equation}}

\def\enq{\end{equation}}

\def\lta{\mathrel{\spose{\lower 3pt\hbox{$\mathchar"218$}}

\raise 2.0pt\hbox{$\mathchar"13C$}}}
\def\gta{\mathrel{\spose{\lower 3pt\hbox{$\mathchar"218$}}
\raise 2.0pt\hbox{$\mathchar"13E$}}}

\def\be{\begin{equation}}
\def\ee{\end{equation}}

\begin{document}

\title{Pulsar Wind Nebulae and the X-Ray Emission of Non-Accreting
Neutron Stars}

\author{K. S. Cheng\altaffilmark{1}, Ronald E. Taam\altaffilmark{2}, and
W. Wang\altaffilmark{1}}

\affil{$^{1}$ Department of Physics, University of Hong Kong,
       Pokfulam Road, Hong Kong\\
     $^{2}$ Northwestern University, Dept. of Physics \& Astronomy,
       2145 Sheridan Rd., Evanston, IL 60208}
\begin{abstract}
The general properties of the non-thermal non-pulsed X-ray
emission of rotation powered pulsars are investigated in the
context of a pulsar wind nebula model. An examination of the
observed X-ray emission from a sample of 23 pulsars in the energy
range between 2-10 keV reveals that the relation of X-ray
luminosity, $L_x$, to the pulsar spin down power, $\dot{E}$, is
steeper for the non-pulsed component than for the pulsed
component.  Specifically, $L_{x}^{npul} \propto \dot{E}^{1.4 \pm
0.1}$ for the non-pulsed component, whereas $L_{x}^{pul} \propto
\dot{E}^{1.2 \pm 0.08}$ for the pulsed component.  The former
relation is consistent with emission from a pulsar wind nebula
model in which $L_{x}^{npul} \propto \dot{E}^{p/2}$ where $p$ is
the power law index of the electron energy distribution.  The
relation for the pulsed component, on the other hand, is
consistent with a magnetospheric emission model. In addition, the
photon spectral index, $\Gamma$, was found to be correlated to the
conversion efficiency of spin down power to non-pulsed X-ray
emission with greater efficiencies for $\Gamma \sim 2 - 2.5$ than
for $\Gamma \sim 1.5 - 2$.  Such a relation is naturally
understood within the framework of a pulsar wind nebula model with
the former relation corresponding to the emission of X-rays in the
fast cooling regime and the latter relation corresponding to
emission in the slow cooling regime.

The X-ray properties of pulsar wind nebulae are sensitive to the
physical conditions (e.g., the density and magnetic field) of the
interstellar medium and can lead to important differences between
the X-ray emission characteristics (luminosity, photon spectral
index and emission morphology) of pulsars in various environments.
Such wind nebulae can contribute to the non-thermal symmetric
emission morphology (point-like) and elongated emission morphology
(tail-like) from sources similar to Geminga and PSR B1757-24.

\end{abstract}


\section{INTRODUCTION}
The study of the emission characteristics from rotation powered
pulsars has been a subject of long standing interest.  Our
knowledge of the fundamental properties (e.g., mass, spin, and
magnetic field) of the underlying neutron star stems from detailed
spectral and timing investigations.  To facilitate an
understanding of the mechanism by which the loss of rotational
energy is converted into high energy radiation, many observational
and theoretical studies have sought to determine the relationship
between the X-ray luminosity, $L_x$, and the rate of rotational
energy loss or spin down power, $\dot{E}$.  Indeed, a correlation
of the form $L_x \propto \dot{E}^{1.39}$ was found in Einstein
data by Seward \& Wang (1988).  Subsequent studies using a larger
sample of pulsars led to a relation of the form, $L_x \propto
\dot{E}$ based on ROSAT data (see Becker \& Tr\"umper 1997) and
$L_x \propto \dot{E}^{1.5}$ based on ASCA data (see Saito 1998).
Recently, a reanalysis of 39 pulsars based on data obtained from
several X-ray satellites by Possenti et al. (2002) led to an
intermediate relation $L_x \propto \dot{E}^{1.34}$, similar to
that of Seward \& Wang (1988).

While the deduced existence of a correlation between $L_x$ and
$\dot{E}$ suggests that the observed X-rays are produced by a
process which taps the rotational energy of the neutron star, a
detailed description of the mechanism remains elusive.  This is,
in part, a result of the fact that the data from different
satellites are obtained in different energy ranges.  The results
can be affected, for example, by interstellar absorption
especially for those pulsars studied in the soft X-ray regime of
ROSAT (0.1- 2.4 keV). In addition, the total X-ray luminosity is
composed of contributions from both the pulsed and non-pulsed
components, and these components are likely to reflect physical
conditions in diverse spatial environments.

The X-rays radiated by rotation powered pulsars could
include contributions from 5 distinct components with different origins: \\
1. Non-thermal non-pulsed diffuse radiation can be emitted from a
pulsar wind nebula.  The radiation will be discussed in \S 2, and
we suggest that this emission component is the major contribution
to the non-pulsed X-ray radiation in the ASCA observations of pulsars.\\
2. Non-thermal non-pulsed radiation from the pulsar magnetosphere
could be important as suggested by Becker et al.  (2004) if the
angular resolution of X-ray detector is sufficiently high to
separate this contribution from that of the nebula. Although such
a component may be present in the ASCA data that we use, due to
its low angular resolution, it is not likely to dominate the
contribution from the pulsar wind nebula. For example, Tennant et
al. (2001) have found a non-thermal non-pulsed component from the
Crab pulsar by using the High-Resolution Camera of the Chandra
X-ray Observatory. They estimate that the X-ray luminosity of this
component is $\sim 10^{34} \rm erg s^{-1}$, whereas, the
non-thermal non-pulsed X-ray luminosity from the Crab nebula
detected by ASCA (Saito et al. 1997a) is $\sim 10^{37} \rm erg
s^{-1}$. The specific origin of the non-thermal non-pulsed
component, not associated with the nebula, is not clear. Although
it contaminates the emission of the nebula, it is small. Provided
that this contribution is small for all the pulsars we study in
this
investigation, this component should not affect our analysis.  \\
3. The non-thermal pulsed radiation component is generally
believed to be produced from the pulsar magnetosphere. This
follows from the fact that the motion of charged particles within
the light cylinder is strongly affected by the magnetic field, and
the radiation of charged particles in the open field lines should
be pulsed. This component could be produced in the vicinity of the
polar cap as a result of inverse Compton scattering of higher
order generation pairs of particles on soft photons emitted by the
neutron star (Zhang \& Harding 2000) or in the outer magnetosphere
as a result of synchrotron radiation of downward cascades from the
outer gap electron/positron particles (Cheng \& Zhang 1999).  \\
4. The thermal non-pulsed radiation from the pulsar surface can
contribute to the soft X-ray bands, typically, characterized by
$kT\sim 0.1$ keV (Cheng \& Zhang 1999). In this paper, we have
used the ASCA data on pulsars so this component will not
contribute to the 2-10 keV ASCA energy band.  \\
5. The thermal pulsed radiation from the pulsar surface likely
originates in a polar cap on the neutron star surface and can
contribute to the hard X-ray band ($kT\sim 1$ keV).  In analyzing
the pulsed emission from pulsars, we follow the model of Cheng \&
Zhang (1999) where the thermal pulsed X-ray emission is included
in their model prediction.

Thus, the X-ray emission observed from pulsars in the
ASCA sample primarily consists of contributions from components 1,
3, and 5. We have assumed that component 2 is not important and
component 4 does not significantly contribute to the 2-10 keV
energy band. On the other hand, Becker \& Tr\"umper(1997) have
used ROSAT data and obtained a linear relation between the X-ray
luminosity consisting of components 2 and 3 and spin-down power.
In the present paper, we concentrate on the X-ray properties in
the energy band of ASCA (2-10 keV). The X-rays,then, consists of a
thermal component from the polar cap, a non-thermal pulsed
component from the magnetosphere
and a non-thermal, non-pulsed emission from the nebula.
That is, we have explicitly assumed that the pulsar magnetosphere
and the pulsar wind nebula are primarily responsible for the
pulsed and non-thermal non-pulsed component respectively.

Since the pulsed and non-pulsed emission have different origins,
the relationship between the total X-ray luminosity and spin down
power is not expected, in general, to be represented by a single
power law. In fact, the non-thermal pulsed and non-pulsed emission
components are distinguished by different spectral signatures and
conversion efficiencies. To determine their relative importance
and their relationship to the spin down power, we have reexamined
the pulsar data obtained in the 2-10 keV energy band of the ASCA
satellite, focusing on separating the non-pulsed and pulsed
emission data. In the next section, we provide a simplified
description of the emission from a pulsar wind nebula, based upon
earlier work by Chevalier (2000), where the
relativistic pulsar wind interacts with the interstellar medium.
In \S 3, the X-ray properties of the pulsed and non-pulsed
components are collected and the relation between their
luminosities with spin down power and photon spectral index are
determined.  With these results in hand, the pulsar wind nebula
model is used to explain the occurrence of the tail-like emission
morphology in Geminga and PSR B 1757-24 in \S 4. Finally, we
summarize and conclude in the last section.

\section{THEORETICAL FRAMEWORK}

A theoretical description of the interaction between a pulsar and
its nebula, as applied to the Crab nebula, was first outlined in a
seminal paper by Rees \& Gunn (1974). In this model, the central
pulsar generates a highly relativistic particle dominated wind
which passes through the medium in the supernova remnant, forming
a shock front. The electrons and positrons in the shock are
envisioned to be accelerated to a power law energy distribution
and to radiate synchrotron photons in the downstream region.

Since the magnetic field in the nebula derives from the pulsar, a
magnetohydrodynamical (MHD) model, rather than a hydrodynamical
model, of the pulsar wind nebula was required. Such a model, in
the steady state and spherically symmetric approximation, was
developed by Kennel \& Coroniti (1984) for the Crab nebula.  A hot
relativistic positronic plasma flow was assumed to be terminated
by a strong MHD shock, decelerating the flow and producing a
non-thermal distribution of electrons.  The magnetization
parameter, $\sigma$, introduced to describe the efficiency for
conversion of energy contained in the pulsar wind into synchrotron
radiation, was represented by the ratio of the electromagnetic
energy flux to the particle kinetic energy flux defined as: \beq
\sigma = {B^2\over 4\pi n\gamma_w mc^2}. \enq Here $n$ is the
particle number density, $B$ is the magnetic field, $\gamma_w$ is
the Lorentz factor of relativistic particles in the wind, $m$ is
the particle mass, and $c$ is the speed of light. As pointed out
by Kennel \& Coroniti (1984), the magnetization parameter must be
small (for the Crab nebula, $\sigma\sim 0.003$) in order that
sufficient compression occurs in the shock for transformation of
the bulk flow energy to random particle motion and to the
subsequent production of the observed synchrotron radiation. Since
the magnetic energy density must dominate the particle energy
density just inside the light cylinder, the wind nebula must have
evolved from a high $\sigma$ to a low $\sigma$ state to produce
the observed properties. The manner in which this transformation
takes place is central for detailed pulsar wind models and is not
understood in detail. Nevertheless, the low $\sigma$ model can
account for the basic high energy properties of the Crab,
including the position of the optical wisps, the spectrum of the
nebula, and the size of the nebula as a function of wavelength
(van den Bergh \& Pritchet 1989).

In the present paper, we concentrate on the X-ray properties of
the pulsar wind nebula, using a simple one zone model similar to
that developed by Chevalier (2000) for a description of the wind
interaction with the surrounding medium in a variety of
environments. Although the model does not provide a description of
the spatial details associated with equatorial tori and polar jets
observed in {\em Chandra} studies of young pulsar wind nebulae, it
can explain the global X-ray properties of the nebula to a first
approximation.

\subsection{One-zone model of X-rays from wind nebulae}

In general, the energy in the shock waves is stored in the
magnetic field as well as in the proton (ion) and electron
particle components.  We assume that the fractional energy density
of the magnetic field, $\epsilon_B$, is $\sim 10^{-3}-10^{-2}$
(Kennel \& Coroniti 1984).  Provided that the spin down power is
eventually carried away by pulsar wind particles and assuming
equipartition of energy between the electrons and protons, the
fractional energy densities in the proton and electron components
should satisfy $\epsilon_p \sim \epsilon_e \sim 0.5$. For a given
$\epsilon_B$ in the shock, the magnetic field at the shock termination
radius, $R_s$ is estimated as $B =(6\epsilon_B \dot{E}/R_s^2
c)^{1/2}$, where $\dot{E}= 4\pi I\dot P /P^3$ is the spin down
power of the pulsar. Here, $P$ and $\dot P$ are the spin period
and its derivative, and $I$ is the moment of inertia of the
neutron star ($\sim 10^{45} {\rm g\ cm^2}$).

At the shock front,
the electrons attain a power law distribution corresponding to
$N(\gamma)\propto \gamma^{-p}$
for $\gamma_m <\gamma <\gamma_{\rm max}$, where $\gamma$ is the
Lorentz factor and $\gamma_m = {p-2\over p-1}\epsilon_e \gamma_w$.
An estimate for $\gamma_ {\rm max}$ can be obtained by equating
the synchrotron cooling timescale to the electron acceleration
timescale.  The former timescale is given by $t_{\rm syn}=6\pi m_e
c/{\sigma_T\gamma B^2}$, and the latter timescale is given by
$t_{\rm acc}= \gamma m_ec/{eB}$, leading to $\gamma_{\max}=(6\pi
e/ {\sigma_TB})^{1/2}$, where $m_e$ is the electron mass and
$\sigma_T$ is the Thompson cross section. The synchrotron power of
an electron with Lorentz factor $\gamma$ is (Blumenthal \& Gould
1970)
\beq
P(\gamma)={4\over 3}\sigma_T c\gamma^2{B^2 \over 8\pi}.
\enq
Following Chevalier (2000), the total rate of particles
ejected by pulsars at the Lorentz factor $\gamma$ is (Chevalier
2000) \beq \dot {N}(\gamma)
=(p-1)\gamma_m^{p-1}(\gamma_wmc^2)^{-1}{\dot{E}} \gamma^{-p}. \enq
The resulting total number of radiating particles becomes
$N(\gamma) \sim \dot {N}(\gamma)t$, where $t$ is a characteristic
timescale.  For a young pulsar with a supernova remnant similar to
the Crab, $t$ is identified with the age of the nebula while for
old pulsars or millisecond pulsars, $t$ is comparable to the flow
timescale within the nebula. Thus, at the emitting frequency of an
electron $\nu(\gamma)=\gamma^2 eB/2\pi m_ec$, the luminosity of
the particle is about $P(\gamma)N(\gamma)$ which should be less
than $\dot{E}$. The detailed properties of the X-ray nebula are a
function of $\epsilon_e$, $\epsilon_B$, $R_s$, $\dot{E}$, and the
two critical frequencies $\nu_m$ and $\nu_c$ where $\nu_m$ is the
frequency radiated by the electrons with the Lorentz factor of
$\gamma_m$, and
\beq
\nu_c=18\pi em_ec/ \sigma_T^2 t^2 B^3
\enq is
the electron synchrotron cooling frequency (see Chevalier 2000 and
references therein).

The X-ray luminosity and spectral index depend on the inequality
between $\nu_X$ and $\nu_c$. For typical values of $\epsilon_e\sim 0.5, \gamma_w\sim 10^6$ and
$B\sim 10^{-4}$ G, $\nu_m\sim 10^{12}$ Hz is always less than
$\nu_X$. In this case, the X-ray luminosity from the total
emission of particles per unit frequency can be calculated
according to the synchrotron spectral profiles which have been
presented in Figure 1 of the paper by Sari, Piran \& Narayan
(1998). For $\nu_X > \nu_c$, the luminosity per unit frequency can
be obtained following equations (7) and (8) of Sari, Piran \&
Narayan (1998)
\beq
L_\nu \simeq (\nu_m/ \nu_c)^{-1/2}
(\nu/\nu_m)^{-p/2}L_{\nu,\rm max},
\enq where $L_{\nu,\rm max}=
N(\gamma_m) P(\gamma_m)$, and $\nu\sim 10^{18}$ Hz. This
expression corresponds to the fast cooling regime, and is
equivalent to equation (9) in Chevalier (2000) reflecting
our use of the same form of particle number injection rate by a pulsar
(see equation 3 in this section). On the other hand, if
$\nu_X < \nu_c$ (denoted as slow cooling), the synchrotron X-ray
luminosity is estimated as (Sari, Piran \& Narayan 1998)
\beq
L_\nu \simeq (\nu/\nu_m)^ {-(p-1)/2}L_{\nu,\rm max}.
\enq
It can
be seen that the X-ray luminosity in the fast cooling regime is
larger than that in the slow cooling regime by a factor
$(\nu_c/\nu_X)^{1/2}$ if $\nu_X$ is in the slow cooling regime.
Since the range of $p$ due to shock acceleration lies between 2
and 3 (see Achterberg et al. 2001; Lemoine \& Pelletier 2003 and
references therein), the photon index in the slow cooling regime
is expected to lie in the range from 1.5 to 2 whereas the photon
index in the fast cooling regime is expected to be in the range of
2 to 2.5.

\subsection{Interactions between the pulsar wind and surrounding medium}

Within the theoretical framework, the energy of the electrons is
converted to X-ray radiation solely at the shock termination
radius.  For pulsar motion which is subsonic, the determination of
this radius is obtained via the balance between the wind ram
pressure and total magnetic and particle pressure within the
nebula (Rees \& Gunn 1974). In this case, the wind bubble will be
nearly centered about the position of the pulsar. Rees \& Gunn
(1974) estimated $R_s \sim 3\times 10^{17}$ cm, which is
consistent with the size of the inner X-ray ring of the Crab
nebula ($\sim 0.1$ pc, see Kennel \& Coroniti 1984; Weisskopf et
al. 2000). The termination shock picture is not necessarily
restricted to slowly moving pulsars since even for pulsars moving
at several hundred km s$^{-1}$ the motion can be subsonic in, for
example, regions where the gas has been heated to temperatures of
$\sim 10^8$ K by a supernova shock. The radius of the termination
shock can be estimated as: \beq R_s \simeq ({\dot{E}\over
B^2c})^{1/2}\sim 6\times 10^{14}\dot{E}_{34}^{1/2}B_{mG}^{-1} {\rm
cm}, \enq where $\dot{E}_{34}$ is the pulsar spin-down power in
units of $10^{34}$ ergs s$^{-1}$ and $B_{\rm mG}$ is the magnetic
field strength in the nebula in milligauss. With the observed
values of the Crab pulsar and its nebula ($\dot{E}_{34} = 5 \times
10^4$ and $B_{\rm mG} = 0.5$), consistency of the shock radius is
easily achieved.

For supersonic motion, the nebula will form a bow shock
morphology. In this case, the termination shock radius is given by
the balance of the ram pressure between the wind particles and the
medium at the head of the bow shock: \beq {\dot{E} \over 4\pi
cR_s^2}={1\over 2}\rho_{\rm ISM} v_p^2, \enq where $\rho_{\rm
ISM}$ is the density of the interstellar medium and $v_p$ is the
pulsar's proper motion velocity. The termination shock radius is
given as: \beq R_s \simeq ({\dot{E}\over 2\pi \rho_{\rm ISM}
v_p^2c})^{1/2}\sim 3\times
10^{16}\dot{E}_{34}^{1/2}n_1^{-1/2}v_{\rm p,100}^{-1} {\rm cm},
\enq where $n_1$, and $v_{\rm p,100}$ are the number density in
the interstellar medium in units of 1 particle cm$^{-3}$ and the
pulsar space velocity in units of 100 km s$^{-1}$.  As an example,
Caraveo et al. (2003) recently discovered a bow shock structure
coincident with the Geminga pulsar based on observations obtained
with {\em XMM-Newton}. An estimate of the termination radius for
Geminga follows from its spin down power, $\dot{E} \simeq
3.2\times 10^{34}{\rm erg\ s^{-1}}$, distance of $\sim 160$ pc,
and proper motion velocity of $v_p\sim 120$ km\ s$^{-1}$ (Bignami
\& Caraveo 1993), leading to a radius of $4\times 10^{16}$ cm
which is consistent with the observational constraint on Geminga's
compact X-ray nebula.

The general features of the emission region surrounding rotation
powered pulsars in the Galaxy will be bracketed by these two wind
nebula morphologies. Since the range in magnetic field strengths
in the shock can vary from $\sim 10^{-5}$ G to as high as
$10^{-4}$ G in the Vela nebula and $5 \times 10^{-4}$ G in the
Crab nebula for the very young pulsars within supernova remnants,
a range in the conversion efficiency of the spin down energy into
nebula luminosity is expected. The relation between the nebula
X-ray luminosity and the pulsar's spin down power will be deferred
until \S 3.

\subsection{General properties of the wind nebulae in X-rays}

In the above we have discussed the pulsar wind nebula models in
different contexts within a one-zone approximation. The
luminosities and spectral properties of the nebulae can be
determined by Eqs. (5) and (6) in the fast and slow cooling regime
respectively. In the fast cooling case, the efficiency of
converting spin down power to nebula luminosity is higher than
that in the slow cooling case. Assuming that $\epsilon_e\sim 0.5,
\epsilon_B \sim 0.01$, $\gamma_w\sim 10^6$ and $p\sim 2.2$ (see
Bednarz \& Ostrowski 1998), the ratio of the X-ray luminosity,
$L_x \sim \nu L_\nu$, to the spin down power at $\nu\sim 10^{18}$
Hz can be estimated as $L_x /\dot{E} \sim 10^{-2}-10^{-1}$ for the
fast cooling case and $L_x/\dot{E} \sim 10^{-4} -10^{-3}$ for the
slow cooling case if $t\sim 10^9$ s, and $B\sim 10^{-5}$ G. The
spectral properties in these two regimes differ since the photon
index in X-rays is $\Gamma=(p+2)/2$, and $\Gamma=(p+1)/2$ for fast
and slow cooling respectively. Since $p$ varies from 2 to 3,
$\Gamma\sim 1.5-2.0$ (slow cooling) and $\Gamma\sim 2.0-2.5$ (fast
cooling). These are typical of the photon indices of observed
X-ray nebulae (see Table 1 of Chevalier 2000).

As described in \S 2.1, the fast and slow cooling regimes are
distinguished by the X-ray
and cooling frequencies.  Since the cooling frequency is dependent
on the magnetic field and the flow time scale of the nebula,
it is possible that different regions of the nebula may
correspond to different cooling regimes. For example, the inner region of
the nebula may correspond to the slow cooling regime, whereas the
fast cooling may be more appropriate in the outer region  of the
nebula. In this case, the power law index of the X-ray spectrum
will steepen by $\sim 0.5$ between the inner and outer region.
This may be consistent with the spectral results from the high
spatial resolution observations of the nebulae surrounding PSR
B1823-13 obtained with the {\em XMM-Newton} satellite (Gaensler et
al. 2003a).

Because the spatial resolution of current X-ray detectors is
limited the emission from compact nebulae of size $\sim R_s$ may
be contaminated by the non-thermal X-ray emission from the pulsar
magnetosphere. However, the pulsar magnetospheric radiation is assumed to
be pulsed, whereas the nebula contribution is non-pulsed.
Therefore, the separation of the non-thermal X-ray emission into
its pulsed and non-pulsed components is essential for determining
their relative contributions.

\section{PULSED AND NON-PULSED EMISSION}

Since the pulsed and non-pulsed emission components originate from
different regions, their X-ray luminosities are not expected to
exhibit a similar relationship to the spin down power.  For
example, Wang \& Zhao (2004) recently found that the correlation
of the observed pulsed X-ray luminosity with spin down power is
not well matched to the linear form found by Becker \& Tr\"umper
(1997) for the total luminosity in the ROSAT energy band (0.1-2.4
keV, see Figure 8 of Wang \& Zhao 2004). We note, however, that
studies in this lower energy band are incomplete since pulsars can
emit a significant fraction of their luminosity at higher
energies. To examine the X-ray luminosity- spin down power
relation at higher energies (2-10 keV), Kawai et al. (1998)
selected a sample of bright nebulae and found a relation in the
form $L_{\rm x} \propto \dot{E}^{1.27 \pm 0.17}$. Since the
individual components were not separately analyzed, the result was
insufficient to establish the relation between the pulsed or
non-pulsed emission with spin down power.

\subsection{$L_x - \dot{E}$ relations from ASCA data}

Over 50 rotation powered pulsars have been detected in the X-ray
band (Becker \& Aschenbach 2002), but only a fraction of them have
been resolved with pulsed non-thermal components.  Here, we have
selected 23 X-ray pulsars with both pulsed and non-pulsed emission
measurements obtained from the ASCA mission (listed in Table 1).
The pulsed and non-pulsed X-ray luminosities of these selected
pulsars are directly taken from the cited references. This pulsar
sample includes 19 normal pulsars and 4 millisecond pulsars whose
observational data are taken from the recent references (see Table
1). Since the ASCA satellite does not have high spatial
resolution, the X-ray luminosity of the pulsars in the ASCA field
is composed of emission from the pulsar's magnetosphere and
compact pulsar wind nebula.  For reference, the total pulsed plus
non-pulsed X-ray luminosity in the ASCA energy range (2-10 keV) is
plotted versus spin down power in Figure 1. A correlation is found
which is consistent with the form $L_x \propto \dot{E}^{1.5}$
found by Saito (1998), but the best fit form of this correlation
is found to be $L_x \propto \dot{E}^{1.35\pm 0.2}$. Here, the
error in the power law exponent represents $\pm 1 \sigma$
corresponding to the scatter in the observed data points, which
may reflect variations in $\epsilon_e$, $\epsilon_B$, $\gamma_w$,
and uncertainties in distance from pulsar to pulsar. Our best fit
power law relation is consistent with the conclusion of Possenti
et al. (2002), who used a sample of 39 pulsars observed mainly by
ROSAT and data from ASCA, RXTE, BeppoSAX, Chandra and XMM-Newton.
Although Possenti et al. (2002) considered their result ($L_x
\propto \dot{E}^{1.34}$) statistically unacceptable, our result is
not subject to the uncertainties associated with the normalization
of different satellite data to obtain the X-ray luminosity between
2-10 keV (cf. Table 1 of Possenti et al. 2002), for which the
extrapolation relied on the uncertain photon index in the ROSAT
energy band.

The X-ray luminosity associated with the pulsed emission component
is illustrated versus spin down power in Figure 2. A correlation
separate from the total luminosity is found which is inconsistent
with either the form $L_{\rm x}\propto \dot{E}$ or $L_x \propto
\dot{E}^{3/2}$. The best fitting function to the data is found to
be $L_{\rm X,pul} \simeq (1.0\pm 0.6) \times 10^{-11}\dot{E}^{1.2
\pm 0.08}$, which significantly deviates from the $3/2$ power law
relation proposed by Saito (1998).   Such a relation is consistent
with the relation $L_x \propto \dot{E}^{1.15}$ derived from the
theoretical X-ray magnetospheric emission model of Cheng \& Zhang
(1999), however, this latter result is not without uncertainties
since the inclination angle of the magnetic field with respect to
the rotation axis and the viewing angle are not well determined.
The observed conversion efficiencies are found to range from $\sim
10^{-5} - 9 \times 10^{-3}$, which is not in conflict with model
predictions (Cheng, Gil \& Zhang 1998; Cheng \& Zhang 1999).

A correlation is also found to exist between the non-pulsed X-ray
luminosity and the spin down power as shown in Figure 3. The data
points are consistent with the previous ASCA relation of the type
$L_{\rm x}\propto \dot{E}^{3/2}$, but the correlation is also
consistent with $L_{\rm x,npul}\propto \dot{E}^{1.4\pm 0.1}$. Upon
comparison to the results from Fig. 1, this power law relation is
a consequence of the fact that the pulsar emission in the ASCA
sample is dominated by the non-pulsed radiation component. The
conversion efficiency for the non-pulsed component overlaps with
that for the pulsed component, but extends to efficiencies as high
as 0.1.

Generally, the X-ray luminosity of pulsar wind nebulae (Chevalier
2000) is a nonlinear function of the spin down power. As can be
seen from equation (9) of Chevalier (2000), the nebula's X-ray
luminosity follows from \beq L_x \propto \epsilon_e^{p-1}
\epsilon_B^{(p-2)/4}\gamma_w^{p-2}R_s^{-(p-2)/2}
\dot{E}^{(p+2)/4}. \enq Here, $\epsilon_e$ and $\epsilon_B$ are
assumed to be constant, but $R_s \propto \dot{E}^{1/2}$ from Eqs.
(7) and (9). Although the above equation corresponds to the fast
cooling regime, the dependence of $L_{X}$ on $\dot{E}$ for the
slow cooling regime is unchanged because the ratio of the X-ray
luminosities in these two regimes depends only on the cooling
frequency $\nu_c$, which is independent of $\dot{E}$.  We note
that $\gamma_w$ also depends on the spin down power so the
explicit dependence of $L_x$ on $\dot{E}$ remains to be
theoretically determined. To estimate this dependence we make use
of the results of Ruderman (1981) and Arons (1983), who argued
that large fluxes of protons (ions) could also be extracted from
the neutron star and accelerated in the parallel electric field in
the magnetosphere. The initial Poynting flux can be converted into
particle thermal and kinetic energy well within the termination
radius. Since both electrons and protons are basically accelerated
by the low frequency electromagnetic wave generated by the pulsar,
they will be accelerated to the same relativistic speed as they
are bound by the strong electrostatic force. Hence, the Lorentz
factor of the electrons and protons are the same, leading to the
result that the protons may have carried away most of the spin
down power (Coroniti 1990).  Thus, we can obtain a form for the
spin down power from \beq \dot{E}\sim \dot N \gamma_wm_pc^2, \enq
where $\dot N$ is the outflow current from the surface.  This
current should be of the order of the Goldreich-Julian current
(Goldreich \& Julian 1969) given as $\dot N \simeq 1.35\times
10^{30}B_{12}P^{-2}{\rm s^{-1}}$. Since $\dot{E}\simeq
10^{31}B_{12}^2 P^{-4}{\rm erg\ s^{-1}}$, we find $\dot N \propto
\dot{E}^{1/2}$, leading to $\gamma_w \propto \dot{E}^{1/2}$.
Therefore, we obtain the relation $L_x\propto \dot{E}^{p/2}$,
where $p$ generally varies between 2 and 3. The relation deduced
from the non-pulsed X-ray luminosity and spin down power of
observed pulsars may result from a relatively high electron energy
index in the nebula.

\subsection{Statistical properties of pulsed and non-pulsed spectra}

The spectral observations of the X-ray emission provide additional
evidence in support of the pulsar wind nebula interpretation. In
particular, the photon indices of both the pulsed and non-pulsed
emission components, where determined, for X-ray pulsars are
listed in Table 2. In this data compilation, the observational
results are taken from the ASCA data and also from the recent
observations by {\em BeppoSAX, Chandra} and {\em XMM-Newton}. In
Figure 4, the photon index of the pulsed component is plotted
versus the ratio of the isotropic pulsed X-ray luminosity to
pulsar spin down power, $L_{\rm pul}/\dot{E}$. The results reveal
that the photon index of the pulsed emission component generally
varies from 1.1 - 1.9 (except for PSR 0218+4232). Such a range is
consistent with the pulsed component originating in the pulsar's
magnetosphere (Wang et al. 1998; Cheng, Gil, \& Zhang 1998). The
values of $L_{\rm pul}/\dot E$ vary over a wide range from
$10^{-5}$ to $10^{-2}$ with no apparent correlation of the photon
index with respect to efficiency. We note that the actual pulsed
X-ray luminosities depend on the beaming of the pulse and,
therefore, the efficiency of spin down power to pulsed X-ray
luminosity is uncertain. The real efficiency may only scatter
around an apparent efficiency, which is based on an assumed
constant solid angle. Unless the beaming conspires with the photon
index to produce a correlation, the rough distribution of the data
presented in Figure 4 would not be expected to change. In summary,
the efficiency and the photon index do not appear to show any
correlation, which is consistent with the model predictions of
Cheng \& Zhang (1999). On the other hand, Cheng \& Zhang (1999)
suggest that the photon index sensitively depends on the local
properties of the stellar magnetic field.

We also illustrate the relation between the efficiency of pulsar
spin down power to non-pulsed X-ray luminosity, $L_{\rm
npul}/\dot{E}$, and the photon index of the non-pulsed component
in Figure 5. It can be seen that the photon index of the
non-pulsed emission component shows two distinct regimes. In
particular, there is a grouping of data points with $\Gamma_{\rm
npul} \lesssim 2$ for low efficiencies ($\lesssim 10^{-3}$) and a
separate grouping of data points with $\Gamma_{\rm npul} \gtrsim
2$ at higher efficiencies. This tendency may reflect a larger
photon index in the fast cooling regime in comparison to the slow
cooling regime as discussed in \S 2.3.  Specifically, we showed
that the fast cooling regime and slow cooling regime is separated
by $(\nu_c/\nu_X) < 1$ or $> 1$ where $\nu_c$ is the cooling
frequency and $\nu_X$ is the observed X-ray frequency
respectively. In the former regime, the photon indices are in the
range from 2-2.5 and in the latter regime, the  photon indices
range from 1.5 to 2. Typically, the slow cooling region has a
lower efficiency than the fast cooling region by a factor
$(\nu_c/\nu_X)^{1/2}<<1$. The actual efficiency of an individual
pulsar depends on its corresponding parameters (e.g. $p, \gamma_w,
B, \epsilon_e$).  Evidence in favor of such a relation is
suggested by the difference between the Crab,  which is in the
fast cooling regime with an efficiency of $\sim 0.03$ and Geminga,
which is in the slow cooling regime with an efficiency of $\sim
10^{-5}$.

\subsection{The termination radius and non-pulsed luminosity}

In Table 3, the termination radii derived from the images of some
identified pulsar wind nebulae are listed. Although the
determination of the radius is imprecise, we have adopted two
procedures for its estimation. For the nebulae which have ring
structures (e.g., Crab nebula), the termination radius is chosen
as the size of the inner ring.  On the other hand, for the
extended sources with a bright point-like central sources, the
termination radius is taken to be the size corresponding to the
scale in which about 90\% of the observed counts are included. For
comparison, the theoretical termination radii for some nebulae are
estimated according to Eqs. (7) (for the Crab nebula) and (9)
(other nebulae) as discussed in \S 2.2 and also listed in Table 3.

With the knowledge of the termination radius of the nebulae, the
X-ray luminosity can be estimated assuming typical values for the
parameters ($\gamma_w$, $p$, $\epsilon_e$, $\epsilon_B$). The
observed and estimated X-ray luminosities of these compact nebulae
are given in Table 3. For these estimates, we have taken $R_s$ as
the observed values, and assumed $\gamma_w\sim 10^{6},\ p=2.2,\
\epsilon_e\sim 0.5$, and $\epsilon_B\sim 0.01$ for all pulsars. We
find that the model luminosities and the observed luminosities
are in rough agreement with each other to within a factor of 4, which taking into
account the possible variations of these parameters amongst the
pulsars, suggests that the simple one-zone model provides a
reasonable approximation for the X-ray luminosity level.

\section {Normal Pulsars and X-ray Tails}

The X-ray emitting region with a characteristic frequency ($\nu_X
=\frac{3 \gamma^2 eB}{2m_ec}$) may exhibit a tail-like spatial
structure provided that the pulsar velocity exceeds that of the
termination shock front and the nebula magnetic field is
sufficiently low. In this case, the distance traversed by the
pulsar within the synchrotron cooling timescale can be taken as a
lower limit of the elongation length.  Specifically, the
synchrotron cooling time in the X-ray band is $\tau_c=6\pi
m_ec/\gamma \sigma_T B^2 \sim 10^8 B_{mG}^{-3/2} (h\nu_X/{\rm
keV})^{-1/2}$ s where $B_{mG}$ is the magnetic field in the
emission region in milligauss. Thus, the typical cooling time is
$\sim 10^{11}$ s for $B_{mG}=0.01$, and the length of the X-ray
elongated feature is about $l\sim v_p \tau_c \sim 10^{18}$ cm for
a pulsar moving at a velocity of 100 km s$^{-1}$ with respect to
the interstellar medium.

The X-ray images of some pulsar wind nebulae, indeed, reveal
extensive X-ray tails. In Table 3, several sources with X-ray tail
features are included. Seven cases are thought to be the
consequence of a bow shock formed by the high velocity pulsar
(e.g., Geminga, PSR B1823-13, B1757-24, B1957-20, J1747-2958,
J1124-5916 and B1853+01). The two pulsars J1930+1852 and B0453-685
exhibit elongated structures and may also be bow shock structures.
On the other hand, the Crab and Vela nebulae exhibit an outflow
structure (X-ray jet).

\subsection{Geminga}

Geminga is one of the most studied pulsars because it is one of
the very nearby pulsars and is the first pulsar detected in
optical, X-ray and gamma-ray, but not in radio (Bignami \& Caraveo
1996) wavelengths. Based on ASCA observations, Saito (1997a) found
that the non-pulsed X-ray luminosity was $\sim 3.6 \times 10^{29}
\rm erg$ s$^{-1}$ and the pulsed X-ray luminosity was $\sim 4.6
\times 10^{29} \rm erg$ s$^{-1}$ in the 2-10 keV band. Based on
more recent XMM-Newton observations, Caraveo et al. (2004) has
reported the detection of a power law component ($\sim 7.7 \times
10^{29} \rm erg$ s$^{-1}$ in 2-8 keV band), which consists of the
pulsed as well as the non-pulsed component. The pulsed component
in 2-6 keV is $\sim (33.2 \pm 4.5)\%$, which is consistent with
ASCA observation. However, it is still not clear if the non-pulsed
component originates from the pulsar or the nebula. The recent
{\em Chandra} observations of the Geminga pulsar show a X-ray bow
shock structure(Caraveo et al. 2003). The image of the field
including the Geminga pulsar reveals a compact point-like X-ray
source of size 20" or ($6\times 10^{16}$ cm) with two elongated
tails extending to 2' (Caraveo et al. 2003). Since the X-ray tails
have a non-thermal X-ray spectrum ($\Gamma\sim 1.6$), these
features may be produced by synchrotron emission in the bow shock
between the pulsar wind and the interstellar medium. The spin down
power of the Geminga pulsar is relatively low compared to other
pulsars, with $\dot{E} \simeq 3.2\times 10^{34} {\rm erg\
s^{-1}}$. Assuming a proper motion velocity of $v_p\sim$120 km\
s$^{-1}$ (Bignami \& Caraveo 1993) and a number density of 1
particle cm$^{-3}$ in the interstellar medium, the termination
shock radius of the pulsar wind nebula is $\sim 4\times 10^{16}$
cm.  This scale is consistent with the observational constraint on
the compact X-ray nebula for an assumed distance of $\sim 160$ pc.
The point-like source has a X-ray luminosity of $1.2\times
10^{30}{\rm ergs\ s^{-1}}$ in the energy range from 0.3 to 5 keV,
and the total luminosity from the two tails is $\sim 6.5 \times
10^{28}$ ergs s$^{-1}$ (Caraveo et al. 2003). The pulsed and
non-pulsed emission of the point-like source was not resolved in
the {\em XMM-Newton} observations, but the results from \S 3 (see
Table 1), reveal that the non-pulsed luminosity of Geminga is
$3.6\times 10^{29}{\rm erg\ s^{-1}}$. Interpreting this luminosity
as arising from the compact X-ray nebula, we find that $L_{\rm
neb,x}/\dot{E} \sim 10^{-5}$ which is low compared to other pulsar
wind nebula. We suggest that the nebula is in the slow cooling
regime where the photon spectral index, $\Gamma=(p+1)/2$.  The
observed photon index of the tail is 1.6, corresponding to
$p=2.2$, and its observed length ($\sim 4\times 10^{17}$ cm)
suggests that the magnetic field of the surrounding medium is weak
($B \sim 1.5 \times 10^{-5}$ G, see also Caraveo et al. 2003).
Thus, the inferred fractional magnetic energy density is
$\epsilon_B\sim 10^{-3}$. Assuming a fractional electron energy
density of $\epsilon_e \sim 0.5$ yields a nebula luminosity of
$\sim 10^{30} {\rm erg\ s^{-1}}$ in the 2-10 keV band, which
modestly overestimates the contributions from the compact nebula
and X-ray tails by a factor of $\sim 2.5$.

\subsection{PSR B1757-24}

Another example where the model can be compared to the
observational data is the X-ray tail associated with PSR B1757-24
observed by Kaspi et al. (2001) with the {\em Chandra} satellite.
PSR B1757-24 is a 124 ms radio pulsar discovered near the
supernova remnant (SNR) G5.4-1.2 (Manchester et al. 1985) with a
spin down power $\dot{E}\sim 2.6\times 10^{36}{\rm erg\ s^{-1}}$
(Manchester et al. 1991).  Assuming an association between the
pulsar and G5.4-1.2, and that the pulsar's characteristic age is a
good estimate of its true age,  a proper motion of 75 mas
yr$^{-1}$ is inferred. This corresponds to a transverse space
velocity of $v_p\sim 1800$ km s$^{-1}$ for a distance of 5 kpc
(Frail, Kassim, \& Weiler 1994). However, recent interferometric
observations have failed to detect the implied proper motion
(Gaensler \& Frail 2000), which suggests that the pulsar is older
than its characteristic age or that the assumed pulsar birth place
is incorrect. Here, we take the proper motion velocity of PSR
B1757-24 to be $v_p\sim 590$ km s$^{-1}$ (Gaensler \& Frail 2000).
The observed 2-10 keV luminosity is about $2\times 10^{33}{\rm
erg\ s^{-1}}$ corresponding to a conversion efficiency of
$10^{-3}$, with the X-ray tail extending to $\sim 20"$ (0.5 pc for
the distance of 5 kpc). The observed photon index, $\Gamma \sim
1.6$, yielding $p=2.2$ in the slow cooling regime. Assuming an
interstellar medium  number density, $n\sim 1$ cm$^{-3}$, the
termination shock radius is $R_s\sim 10^{17}$ cm, with the known
velocity $v_p\sim 590$ km s$^{-1}$. Taking the typical pulsar wind
nebula parameters, $\epsilon_e\sim 0.5,\ \epsilon_B\sim 0.01$
(corresponding to $B\sim 20 \mu$G at $R_s$), and $\gamma_w \sim
10^6$, we find the X-ray luminosity from 2-10 keV is $\sim 3\times
10^{33} {\rm erg\ s^{-1}}$ using the one-zone model. The predicted
X-ray tail length is given by $l\sim v_p t_c\sim 2\times 10^{18}$
cm. Both these estimates for the X-ray luminosity and tail length
are in approximate accord with the observed values.

\section{SUMMARY}

The non-thermal non-pulsed X-ray emission of rotation powered
pulsars has been investigated in the context of emission from a
pulsar wind nebula. We have confirmed that this emission can
significantly contribute to the total X-ray emission, thereby,
increasing the X-ray emission above that produced in the pulsar
magnetosphere.  A reexamination of ASCA data for pulsars separated
into the non-pulsed and pulsed emission components indicates that
their non-thermal components, individually, are correlated with
spin down power.  In particular, the dependence is steeper for the
non-pulsed component with $L_x \propto \dot{E}^{1.4\pm 0.1}$ as
compared to the relation, $L_x \propto \dot{E}^{1.2\pm 0.08}$ for
the pulsed component. Our results are similar to that discovered
by Becker \& Tr\"umper(1997), who found that the non-thermal
(pulsed and non-pulsed) component satisfied the linear relation,
$L_x \propto \dot{E}$, based on ROSAT data in 0.1-2.4 keV range.
In this paper, we have used the ASCA data in the 2-10 keV range,
which includes the contribution of the additional thermal pulsed
component with $kT \sim 1$ keV from the polar cap (Cheng \& Zhang
1999). In this case, the dependence of the X-ray luminosity on
spin down power is steeper.   Because of the poor angular
resolution of ASCA, the possible contribution of non-thermal
non-pulsed emission from the magnetosphere is mixed in with that
of the nebula.

Within the framework of a one zone model for the pulsar wind
developed by Chevalier (2000), the power law relation between the
non-pulsed luminosity and spin down power provides an important
diagnostic for the electron energy distribution in the shock
resulting from the interaction of the relativistic wind with the
interstellar medium. Specifically, $L_x \propto \dot{E}^{p/2}$
where $p$ is the power law index of the electron energy
distribution, suggesting that the index characterizing these
pulsars is relatively high, $p \sim 2-3$.  The observational
results also show that the conversion efficiency of spin down
power to X-ray luminosity is a function of spin down power varying
from $10^{-5}$ to 0.1.  The efficiencies in the lower (upper) end
of the range can be understood in terms of emission in the slow
(fast) cooling regime where the X-ray frequency is less (greater)
than the electron synchrotron cooling frequency.

The spectra from pulsar wind nebulae are distinctly non-thermal,
described by a photon power law index, $\Gamma = (p+2)/2$ or
$\Gamma = (p+1)/2$ for the fast or slow cooling regime
respectively. This implies a correlation between the photon index
and efficiency of conversion of spin down power to non-pulsed
X-ray luminosity with steeper photon power law indices associated
with higher conversion efficiencies.  Although the photon power
law indices inferred from the pulsars studied have large
uncertainties, there are hints from the observed data that such a
correlation exists.  We note that the low values of $\Gamma$ and
$L_x/\dot{E}$ inferred for the extended emission observed from
Geminga and B 1757-24 are consistent with such an interpretation.
Future observed determinations of the photon power law indices
could further verify this correlation, thereby, providing an
additional diagnostic tool for probing the pulsar and its
environment.

Finally, the pulsar wind nebula interpretation may also apply to
the spatially extended X-ray emission observed in binary systems
containing a pulsar.  For example, the extended X-ray emission
associated with the millisecond pulsar B1957+20 (Stappers et al.
2003) may be interpreted within such a framework.  The application
of such models to recycled pulsars in the Galaxy will be the
subject of a future investigation.

\acknowledgements We are grateful to the anonymous referee for the
fruitful suggestions. This work is partially supported by the NSF
through grant AST-0200876, by a RGC grant of the Hong Kong
Government, and by the National Natural Science Foundation of
China under grant 10273011.

\clearpage
\begin{figure}
\psfig{figure=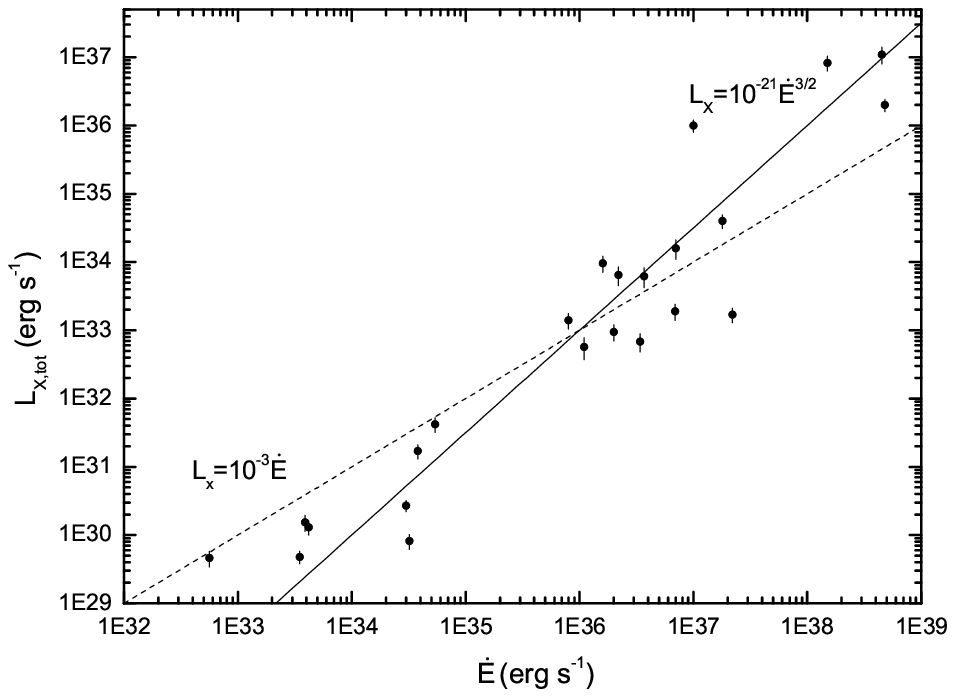,angle=0,width=14cm} \caption{The total X-ray
luminosity (2-10 keV) from ASCA observations versus spin-down
power of 23 X-ray pulsars. The solid line is $L_{\rm
X}=10^{-21}\dot{E}^{3/2}$  as used by Saito (1998), and the dashed line
represents $L_{\rm X}=10^{-3}\dot{E}$ as used by Becker \& Tr\"umper (1997)
to summarize ROSAT observations of central objects and pulsars.
The best fitting function is $L_{\rm X} \propto \dot{E}^{1.35\pm 0.2}$, where the
error in the power law exponent represents $\pm 1 \sigma$
corresponding to the scatter in the observed data points.}
\end{figure}

\begin{figure}
\psfig{figure=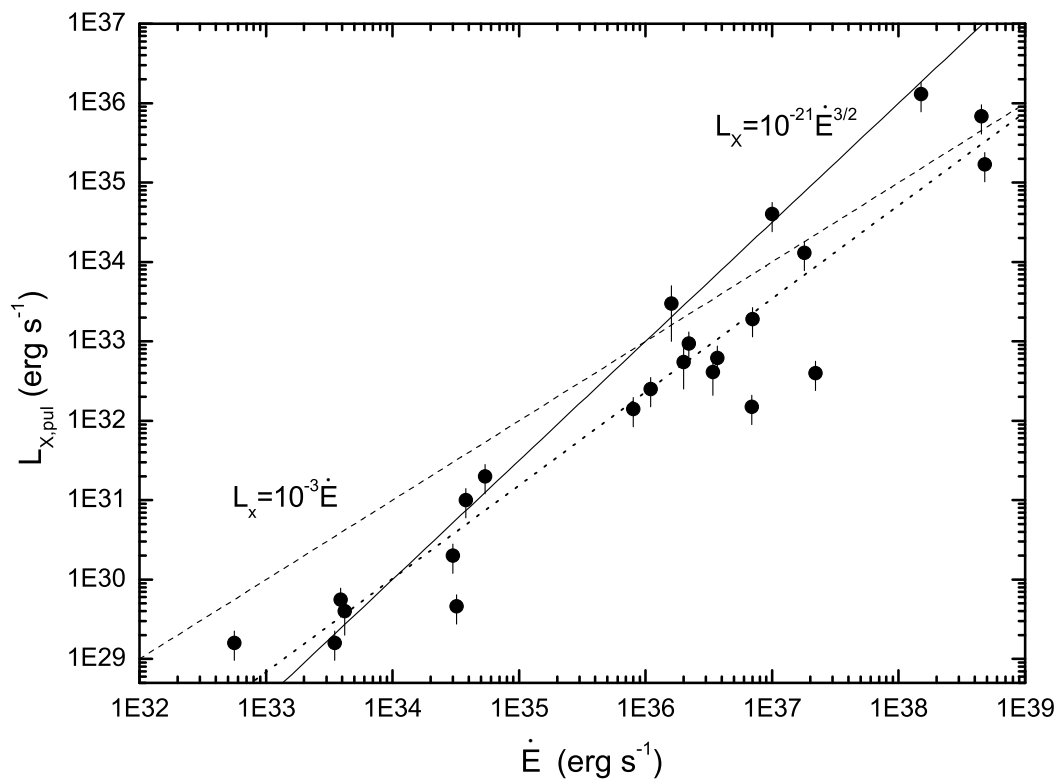,angle=0,width=14cm} \caption{The pulsed X-ray
luminosity (2-10 keV) from ASCA observations versus spin-down
power of 23 X-ray pulsars. The solid line is $L_{\rm
X}=10^{-21}\dot{E}^{3/2}$, and the dashed line represents $L_{\rm
X}=10^{-3}\dot{E}$. The relation between the pulsed component and
spin-down power cannot be described by both the two formulae. The
best fitting function is shown as the dotted line, $L_{\rm
X,pul}=10^{-11}\dot{E}^{1.2\pm 0.08}$. }

\end{figure}

\begin{figure}
\psfig{figure=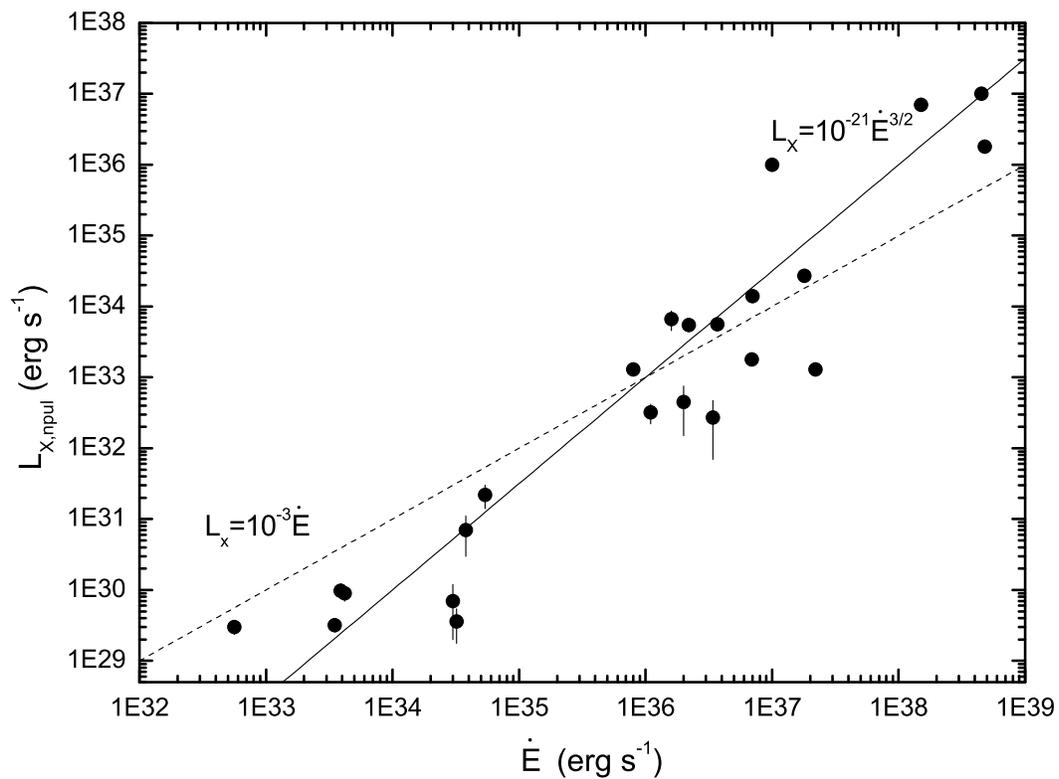,angle=0,width=14cm} \caption{The non-pulsed
X-ray luminosity (2-10 keV) from ASCA observations versus
spin-down power of 23 X-ray pulsars. The solid line is $L_{\rm
X}=10^{-21}\dot{E}^{3/2}$, and the dashed line represents $L_{\rm
X}=10^{-3}\dot{E}$. The best fitting function is $L_{\rm X}
\propto \dot{E}^{1.4\pm 0.1}$.}
\end{figure}

\begin{figure}
\psfig{figure=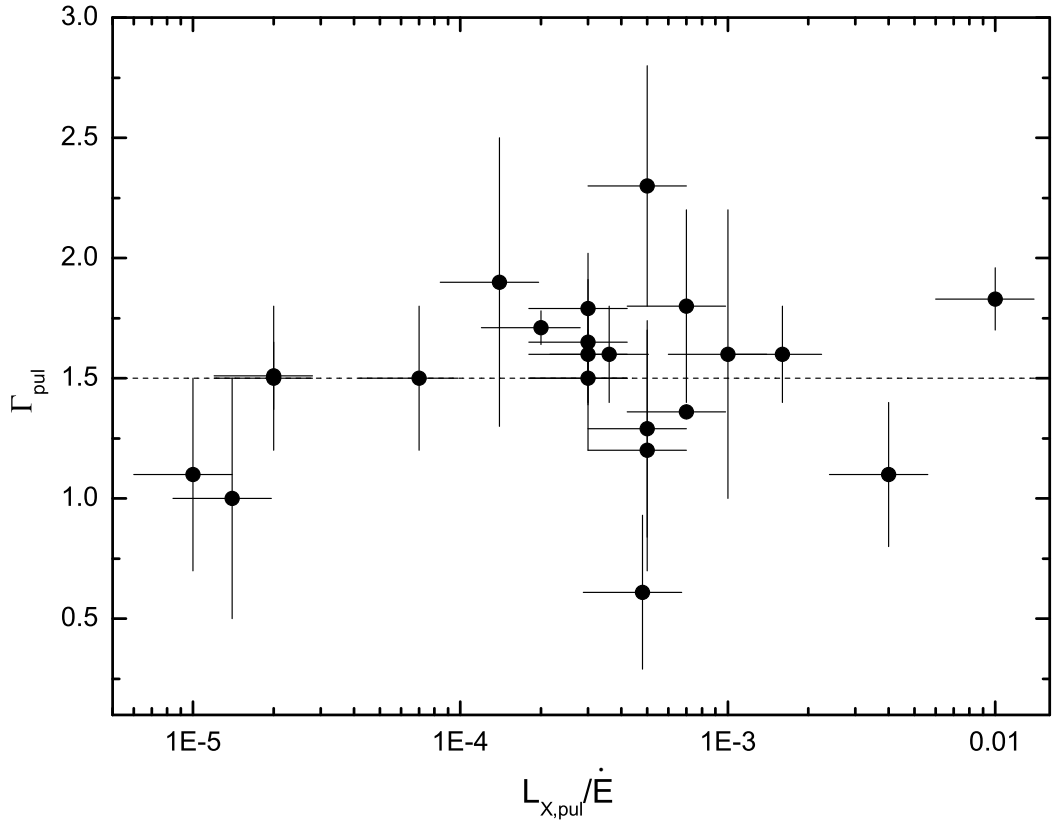,angle=0,width=14cm} \caption{The photon index
of the pulsed emission component of spin powered X-ray pulsars
versus $\eta_{\rm pul}=L_{\rm pul}/\dot{E}$, the ratio of the
isotropic pulsed X-ray luminosity to spin down power. }
\end{figure}

\begin{figure}
\psfig{figure=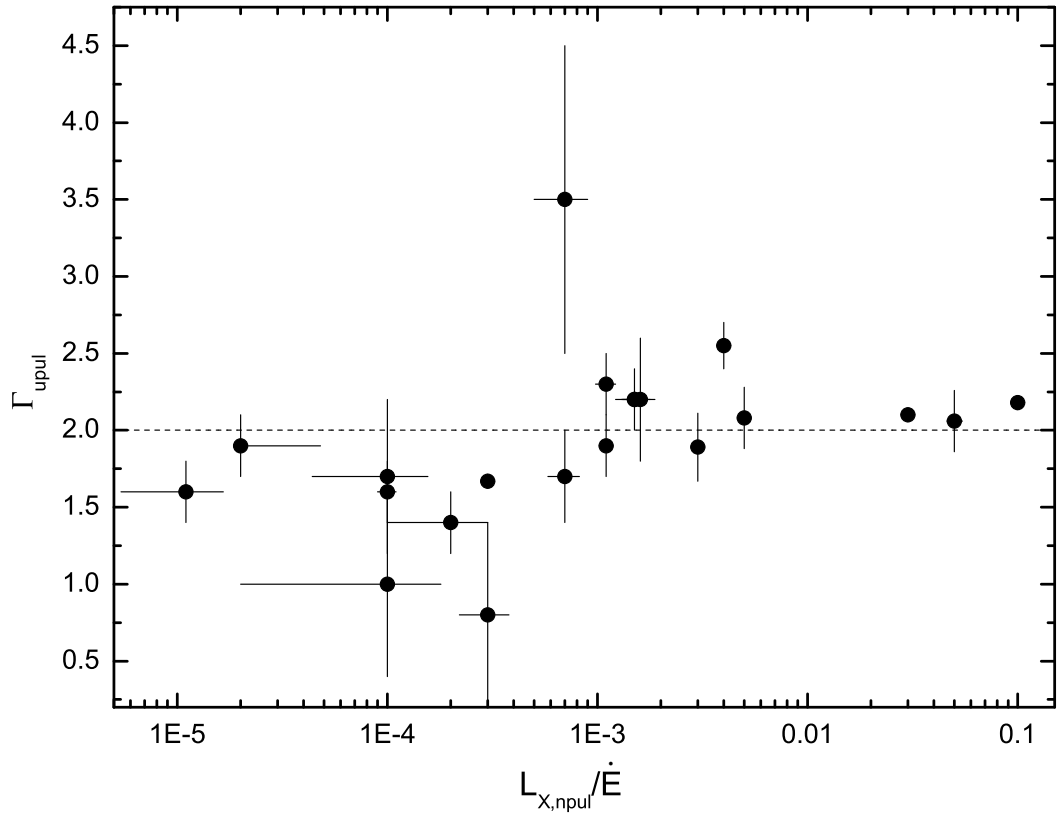,angle=0,width=14cm} \caption{The photon index
of the non-pulsed emission component of spin powered X-ray pulsars
versus $\eta_{\rm npul}=L_{\rm npul}/\dot{E}$, the efficiency of
spin down power to non-pulsed X-ray luminosity. The dashed line
corresponding to $\Gamma=2$, which approximately separates the
fast cooling regime from the slow cooling regime, is shown for
convenience.}

\end{figure}

\clearpage

\begin{table}
\caption{Characteristics of spin powered pulsars and their X-ray
luminosities observed by ASCA}
\begin{center}
\scriptsize
\begin{tabular}{l c c c c c c c l}
\tableline \tableline PSR & $P$ (s) & $\dot P({\rm s\ s^{-1}})$ &
$d$ & $\dot{E}$ & $L_{\rm X,tot}$ & $L_{\rm X,pul}$ &
$L_{\rm X,npul}$ & Reference \\
\tableline J0631+1036 & 0.288 &  $1.0\times 10^{-13}$ &  1.0  &
$5.4\times 10^{34}$ & $4.2\times 10^{31}$ & $2.0\times 10^{31}$ &  $2.2\times 10^{31}$ &  1 \\
J1811-1926 & 0.065  & $4.4\times 10^{-14}$  & 5.0 &  7.0$\times
10^{36}$  & 1.6$\times 10^{34}$ & 1.9$\times 10^{33}$ & 1.4$\times
10^{34}$ &  2,3 \\
B0531+21 &  0.033  & 4.2$\times 10^{-13}$ &  2.0 & 4.5$\times
10^{38}$ & 1.1$\times 10^{37}$ & 6.8$\times 10^{35}$ & 1.0$\times
10^{37}$ &  4 \\
B0833-45 &  0.089 &  1.25$\times 10^{-13}$ & 0.3  & 6.9$\times
10^{36}$ & 1.9$\times 10^{33}$ &  1.5$\times 10^{32}$ & 1.8$\times
10^{33}$ &  4,5,6 \\
B0633+17 &  0.237  & 1.1$\times 10^{-14}$  & 0.16 & 3.2$\times
10^{34}$ & 8.2$\times 10^{29}$ & 4.6$\times 10^{29}$ & 3.6$\times
10^{29}$ &  7,8 \\
B1706-44  & 0.1025 & 9.3$\times 10^{-14}$  & 1.82 & 3.4$\times
10^{36}$ & 6.8$\times 10^{32}$ &  4.1$\times 10^{32}$ & 2.7$\times
10^{32}$ &  4 \\
B1509-58 &  0.150 &  1.54$\times 10^{-12}$ & 4.3  & 1.8$\times 10^{37}$
& 4.0$\times 10^{34}$ &  1.3$\times 10^{34}$ & 2.7$\times 10^{34}$ &  4 \\
B1951+32  & 0.0395 & 5.8$\times 10^{-15}$ &  2.5 &  3.7$\times
10^{36}$ & 6.2$\times 10^{33}$ & 6.2$\times 10^{32}$ & 5.6$\times
10^{33}$ &  4 \\
B1046-58  & 0.124  & 9.6$\times 10^{-14}$  & 2.98 & 2.0$\times
10^{36}$ &  9.5$\times 10^{32}$ & 5.5$\times 10^{32}$ & 4.5$\times 10^{32}$ & 4,9 \\
B1929+10 &  0.227  & 1.16$\times 10^{-15}$ & 0.17 & 3.9$\times
10^{33}$ & 1.54$\times 10^{30}$ & 5.6$\times 10^{29}$ & 9.8$\times 10^{29}$ &  10 \\
B0656+14  & 0.385  & 5.5$\times 10^{-14}$  & 0.76 & 3.8$\times
10^{34}$ & 1.7$\times 10^{31}$ & 1.0$\times 10^{31}$ & 7.0$\times
10^{30}$ &  4 \\
B0540-69  & 0.05  &  4.8$\times 10^{-13}$ &  49.4 & 1.5$\times
10^{38}$ & 8.3$\times 10^{36}$ & 1.3$\times 10^{36}$ & 7.0$\times
10^{36}$ &  11 \\
B0950+08  & 0.253 &  2.3$\times 10^{-16}$ &  0.12 & 5.6$\times
10^{32}$ & 4.6$\times 10^{29}$ & 1.6$\times 10^{29}$ & 3.0$\times
10^{29}$ &  10 \\
B1610-50  & 0.232 &  4.93$\times 10^{-13}$  & 7.26 & 1.6$\times
10^{36}$ &  9.6$\times 10^{33}$ & 3.0$\times 10^{33}$ & 6.6$\times
10^{33}$ &  9 \\
B1055-52  & 0.197 & 5.83$\times 10^{-15}$ & 1.53 & 3.0$\times
10^{34}$ &  2.7$\times 10^{30}$ & 2.0$\times 10^{30}$ & 7.0$\times
10^{29}$ &  12 \\
B1853+01  & 0.267 &  5.4$\times 10^{-13}$ &  2.02 & 8.0$\times
10^{35}$ & 1.4$\times 10^{33}$ & 1.4$\times 10^{32}$ & 1.3$\times
10^{33}$ &  4 \\
J2229+6114 & 0.0516 & 7.8$\times 10^{-14}$ & 3.0 & 2.2$\times
10^{37}$ & 1.7$\times 10^{33}$ & 4.0$\times 10^{32}$ & 1.3$\times
10^{33}$ & 13\\
B0537-69 & 0.016 & 5.13$\times 10^{-14}$ & 47 & 4.8$\times
10^{38}$ & 2$\times 10^{36}$ & 1.7$\times 10^{35}$ & 1.8$\times
10^{36}$ & 14,15\\
J1846-0258 & 0.32 & $7.1\times 10^{-12}$ & 19 & 1$\times 10^{37}$
& 1$\times 10^{36}$ & 4$\times 10^{34}$
& 1$\times 10^{36}$ & 16 \\
B1937+21  & 0.00156 & 1.05$\times 10^{-20}$ & 3.6 & 1.1$\times
10^{36}$ & 5.7$\times 10^{32}$ & 2.5$\times 10^{32}$  & 3.2$\times
10^{32}$ &  17 \\
J2124-3358 & 0.005 &  1.08$\times 10^{-20}$ & 0.25 & 3.5$\times
10^{33}$ & 4.8$\times 10^{29}$ & 1.6$\times 10^{29}$ & 3.2$\times
10^{29}$  &  18 \\
B1821-24 &  0.003 &  1.6$\times 10^{-18}$  & 5.1 &  2.2$\times
10^{36}$ & 6.5$\times 10^{33}$ &  9.4$\times 10^{32}$ & 5.5$\times 10^{33}$ &  19  \\
J0437-47  & 0.0058 & 2.0$\times 10^{-20}$ &  0.18 & 4.2$\times
10^{33}$ & 1.3$\times 10^{30}$ & 4.0$\times 10^{29}$ & 9.0$\times
10^{29}$  & 4   \\
\tableline
\end{tabular}
\end{center}

\tablecomments{The first column PSR is the pulsar name, $P$ is the
spin period, $\dot P$ is the period derivative, $d$ is the
distance of the pulsar from us in units of kpc. The luminosity is
in units of erg\ s$^{-1}$. $\dot{E}$ is the pulsar's spin-down
power. $L_{\rm X,tot}$ is the total X-ray luminosity observed by
ASCA (2-10 keV), $L_{\rm X,pul}$ is just the pulsed X-ray
luminosity, $L_{\rm X,npul}$ is the non-pulsed luminosity. Because
the X-ray luminosities are all calibrated by ASCA, we can assume a
10\% uncertainty for the total flux measurement, and except for
B1706-44, B1046-58, J1610-50 and J0437-47 which have a 50\%
uncertainty for the pulsed fraction, the others have 30\% fraction
uncertainties. \\
References: 1. Torii et al. 2001; 2. Torii et al. 1997; 3. Torii
et al. 1999; 4. Saito et al. 1997a; 5. Helfand, Gotthelf \&
Halpern 2001; 6. Pavlov et al. 2001; 7. Halpern \& Wang 1997; 8.
Cavaveo et al. 2003;  9. Pivovaroff et al. 2000; 10. Wang \&
Halpern 1997; 11. Hirayama et al. 2002;  12. Shibata et al. 1997;
13. Halpern et al. 2001; 14. Marshall et al. 1998; 15. Wang \&
Gotthelf 1998; 16. Gotthelf et al. 2000; 17. Takahashi et al.
2001; 18. Sakurai et al. 2001; 19. Saito et al. 1997b. }
\end{table}

\begin{table}
\caption{X-ray pulsed and non-pulsed component properties of
spin-powered pulsars}
\begin{center}
\begin{tabular}{l c c c c c l}
\tableline \tableline PSR & $\dot{E}$ & $\eta_{\rm pul}$ &
$\eta_{\rm npul}$ & $\Gamma_{\rm pul}$  & $\Gamma_{\rm npul}$  &
Reference \\
\tableline
B1823-13  & 2.8$\times 10^{36}$ & $3\times 10^{-4}$ & $1.1\times
10^{-3}$ & $1.6\pm 0.1$ & $2.3\pm 0.2$ &  1  \\
J0537-6910 & 4.8$\times 10^{38}$ & $3.6\times 10^{-4}$ & $4\times
10^{-3}$ & $1.6\pm 0.2$ & $2.55\pm 0.15$ &  2  \\
B0540-69  & 1.5$\times 10^{38}$  & 0.01 & 0.05 & $1.83\pm
0.13$  & $2.06\pm 0.2$ &  3  \\
B1509-58  & 1.8$\times 10^{37}$ & $7\times 10^{-4}$ & $1.5\times
10^{-3}$  & $1.36\pm
0.02$ & $2.2\pm 0.005$ &  4  \\
J0218+4232 & 2.5$\times 10^{35}$ & 4.8$\times 10^{-4}$ & 2$\times
10^{-4}$ &
0.61$\pm 0.32$ & $1.4\pm 0.2$ &  5 \\
J0631+1036 & 5.4$\times 10^{34}$ & $5\times 10^{-4}$ & $7\times
10^{-4}$ &
2.3$\pm 0.5$ & $3.5\pm 1.0$ &   6  \\
J1811-1926 & $7\times 10^{36}$ & 3$\times 10^{-4}$ & $5\times
10^{-3}$ &
1.65$\pm 0.26$ & $2.08\pm 0.2$ &   7  \\
B0531+21 & 4.5$\times 10^{38}$ & 1.6$\times 10^{-3}$ & 0.03 &
1.6$\pm 0.2$ &
2.1$\pm 0.01$ &  8 \\
B0833-45 & $6.9\times 10^{36}$ & 2$\times 10^{-5}$ & 3$\times
10^{-4}$ &
1.5$\pm 0.3$ & 1.67$\pm 0.04$ &  9 \\
B0633+17 & 3.2$\times 10^{34}$ & 1.4$\times 10^{-5}$ & 1.1$\times
10^{-5}$ &
1.0$\pm 0.5$ & 1.6$\pm 0.2$ &  10,11  \\
B1055-52 & 3$\times 10^{34}$ & 7$\times 10^{-5}$ & 2$\times
10^{-5}$ &
1.5$\pm 0.3$ & 1.9$\pm 0.2$ &    12 \\
B1937+21 & 1.1$\times 10^{36}$ & 2$\times 10^{-4}$ & $3\times
10^{-4}$ &
1.71$\pm 0.07$ & 0.8$\pm 0.6$ &    13  \\
B1821-24 & 2.2$\times 10^{36}$ & 5$\times 10^{-4}$ & 3$\times
10^{-3}$ &
1.2$\pm 0.5$ & $1.89\pm 0.22$ &   14  \\
J2229+6114 & 2.2$\times 10^{37}$ & 2$\times 10^{-5}$ & $10^{-4}$ &
1.51$\pm 0.14$ & 1.6$\pm 0.2$ &  15 \\
J1105-6107 & 2.5$\times 10^{36}$ & 6$\times 10^{-4}$ & 1.6$\times
10^{-3}$ & 1.8$\pm 0.4$ &
2.2$\pm 0.4$ &  16  \\
B1706-44 & 3.4$\times 10^{36}$ & 1.4$\times 10^{-4}$ & $10^{-4}$ &
1.9$\pm 0.6$ & 1.7$\pm 0.5$ &  17 \\
B1757-24 & 2.6$\times 10^{36}$ & $10^{-3}$ & $10^{-4}$ &  1.6$\pm
0.6$ & 1.0$\pm 0.6$ &  18  \\
J0205+6449 & 2.6$\times 10^{37}$ & $10^{-5}$ & $1.1\times 10^{-3}$
& 1.1$\pm 0.4$ & 1.9$\pm 0.2$ &  19 \\
B1929+10 & 3.9$\times 10^{33}$ & 3$\times 10^{-4}$ & $6\times
10^{-4}$ & 1.79$\pm 0.23$ &  &   20  \\
B0656+14 & 3.8$\times 10^{34}$ & 3$\times 10^{-4}$ & 2$\times
10^{-4}$ & 1.5$\pm 0.1$ & &  21 \\
J2021+3651 & $3.6\times 10^{36}$ & 3$\times 10^{-4}$ & $7\times
10^{-4}$ & 1.5$\pm 0.3$ & 1.7$\pm 0.3$ & 22 \\
J1747-2958 & 2.5$\times 10^{36}$ & $10^{-3}$ & 0.02 & 1.8$\pm 0.1$
& 2.1$\pm 0.1$ & 23 \\
B1853+01 & $4.3\times 10^{35}$ &  $5\times 10^{-4}$ & 1.5$\times
10^{-3}$ & $1.29\pm 0.45$  & $2.2\pm 0.2$ & 24 \\
\tableline

\end{tabular}
\end{center}

\tablecomments{The spin down power is in units of erg\ s$^{-1}$.
$\eta_{\rm pul}=L_{\rm pul}/\dot{E}$ and $\eta_{\rm npul}=L_{\rm
npul}/\dot{E}$ are the X-ray efficiencies of spin down power to
pulsed and non-pulsed X-ray luminosities separately. $\Gamma_{\rm
pul}$ and $\Gamma_{\rm npul}$ are the pulsed and
non-pulsed photon indices respectively. \\
References: 1. Gaensler et al. 2003; 2. Marshall et al. 1998; 3.
Kaaret et al. 2001; 4. Marsden et al. 1997, 5. Mineo et al. 2000;
6. Torii et al. 2001; 7. Torii et al. 1999;  8. Willingale et al.
2001; 9. Pavlov et al. 2001; 10. Halpern \& Wang 1997; 11. Cavaveo
et al. 2003; 12. Shibata et al. 1997; 13. Takahashi et al. 2001;
14. Saito et al. 1997b; 15. Halpern et al. 2001. 16. Gotthelf \&
Kaspi 1998; 17. Finley et al. 1998; 18. Kaspi et al. 2001; 19.
Murray et al. 2001; 20. Wang \& Halpern 1997; 21. Greiveldinger et
al. 1996; 22. Hessels et al. 2004; 23. Gaensler et al. 2004; 24.
Petre et al. 2002.}
\end{table}

\begin{table}
\caption{X-ray properties of pulsar wind nebulae}
\begin{center}
\scriptsize
\begin{tabular}{l c c c c c c c c l}
\tableline \tableline PSR & $\dot{E}$ & $R_s^{\rm obs}$ &
$R_s^{\rm th}$ & $L_{\rm upul}^{\rm obs}$ & $L_{\rm upul}^{\rm
th}$ & $n$ & $v_p$ & $l^\dag$ & Reference \\
\ &  & cm & cm & & & cm$^{-3}$ & km\ s$^{-1}$ & cm & \
\\
\tableline B1823-13 & 2.8$\times 10^{36}$ & 6$\times 10^{17}$ & &
3$\times 10^{33}$ & 2$\times 10^{33}$ & 1 &  & 7$\times 10^{18}$ & 1  \\
J0537-6910 & 4.8$\times 10^{38}$ & 4$\times 10^{17}$ & & 1.8$\times 10^{36}$ & 3$\times 10^{36}$ & & & & 2 \\
B0540-69  & 1.5$\times 10^{38}$  &  3$\times 10^{17}$ & & 7$\times 10^{36}$ & 4$\times 10^{36}$ & & & & 3 \\
J1811-1926 & $7\times 10^{36}$ & 2$\times 10^{17}$ & & 1.4$\times 10^{34}$ & $10^{34}$ & & & & 4  \\
B0531+21 & 4.5$\times 10^{38}$ & 4$\times 10^{17}$ & 3$\times
10^{17}$ & $10^{37}$
& 4$\times 10^{36}$ & 10 & 123 & $10^{18}$ & 5 \\
B0833-45 & $6.9\times 10^{36}$ &  $10^{17}$ & 2$\times
10^{17}$ & 1.8$\times 10^{33}$ & $10^{33}$ & 1 & 65 & 8$\times 10^{16}$ & 6 \\
B0633+17 & 3.2$\times 10^{34}$ & 5$\times 10^{16}$ & 4$\times
10^{16}$
& 4$\times 10^{29}$ & $10^{30}$ & 1 & 120 & 4$\times 10^{17}$ & 7 \\
J2229+6114 & 2.2$\times 10^{37}$ & 4$\times 10^{17}$ & & 1.3$\times 10^{33}$ & 3$\times 10^{33}$ & & & & 8  \\
J1105-6107 & 2.5$\times 10^{36}$ & 4$\times 10^{16}$ & & 4$\times 10^{33}$ & 3$\times 10^{33}$ & & & & 9 \\
B1706-44 & 3.4$\times 10^{36}$ & 3$\times 10^{17}$ & & 3$\times 10^{32}$& 4$\times 10^{32}$ & & & & 10 \\
B1757-24 & 2.6$\times 10^{36}$ & 4$\times 10^{16}$ & $5\times
10^{16}$ & 3$\times 10^{32}$ & 5$\times 10^{32}$ & 1 & $590$ & $10^{18}$ & 11 \\
J0205+6449 & 2.6$\times 10^{37}$ & $10^{17}$ & & 3$\times 10^{34}$ & 3$\times 10^{34}$ & & & & 12 \\
B1957+20 & $10^{35}$ & 5$\times 10^{16}$ & 4$\times 10^{16}$
& 1.6$\times 10^{31}$ & $10^{31}$  & 1 & 220 & 4$\times 10^{17}$ & 13 \\
J2021+3651 & $3.6\times 10^{36}$ & 8$\times 10^{17}$ &  & 3$\times
10^{33}$ &  2$\times 10^{33}$ &  &  & & 14 \\
J1747-2958 & 2.5$\times 10^{36}$ & 3$\times 10^{16}$ & 7$\times
10^{16}$ & 5$\times 10^{34}$ & $10^{34}$ & 0.3 & 600 & 2$\times 10^{18}$ & 15,16  \\
J1124-5916 & $10^{37}$ & $10^{17}$ & 2$\times 10^{17}$ & 4$\times
10^{34}$ & $10^{34}$ & 0.5 & 450 & 6$\times 10^{17}$ & 17 \\
B1853+01 & $4.3\times 10^{35}$ & 10$^{17}$ & 3$\times 10^{16}$ &
6$\times 10^{32}$
& 4$\times 10^{32}$ & 5 & 375 & $10^{18}$ & 18 \\
J1930+1852 & 2$\times 10^{36}$ & $10^{17}$ &  & $10^{33}$ &
7$\times 10^{32}$ & 1 & & $10^{18}$ & 19 \\
B0453-685  & $10^{37}$ & 6$\times 10^{17}$ & & 6$\times 10^{34}$ &
2$\times 10^{34}$ & 0.4 & & $3\times 10^{18}$ & 20
\\
J0538+2817 & 4$\times 10^{34}$ & 8$\times 10^{16}$ & 3$\times
10^{16}$ & 6$\times 10^{31}$ & 5$\times 10^{31}$ & 0.5
& 385 & & 21 \\
\tableline
\end{tabular}
\end{center}
\tablecomments{The luminosity is in units of erg\ s$^{-1}$.
$R_s^{\rm obs}$ and $R_s^{\rm th}$ are the observed and predicted
termination radius of the pulsar wind nebulae. $L_{\rm upul}^{\rm
obs}$ and $L_{\rm upul}^{\rm th}$ are the observed and theoretical
X-ray luminosities of pulsar wind nebulae. $n$ is the number
density of the medium around the pulsar. $v_p$ is the pulsar
proper motion velocity, $l$ is the length of the observed
X-ray elongated structure.  \\
$^\dag$ Seven fast moving pulsars in the bow shock show the X-ray
tail features. $l$ of the pulsars Crab and Vela corresponds to the
outflow scales of the nebulae. For B0453-685 and J1930+1852, $l$
how the scales of the elongated structure. \\
References: 1. Gaensler et al. 2003a; 2. Marshall et al. 1998; 3.
Kaaret et al. 2001; 4. Reynonds et al. 1994; 5. Weisskopf et al.
2000; 6. Pavlov et al. 2001; 7. Cavaveo et al. 2003; 8. Halpern et
al. 2001. 9. Gotthelf \& Kaspi 1998; 10. Finley et al. 1998; 11.
Kaspi et al. 2001; 12. Murray et al. 2001; 13. Stappers et al.
2003; 14. Hessels et al. 2004; 15. Gaensler et al. 2004; 16.
Camilo et al. 2002; 17. Hughes et al. 2001; 18. Petre et al. 2002;
19. Lu et al 2002; 20. Gaensler et al. 2003b; 21. Romeni \& Ng
2003.}
\end{table}

\end{document}